\documentclass[letterpaper, 10 pt, conference]{ieeeconf}  

\IEEEoverridecommandlockouts

\usepackage{graphicx} 
\usepackage{amsmath,amssymb,amsfonts}
\usepackage{physics}
\usepackage[most]{tcolorbox}
\usepackage{xcolor}
\usepackage{colortbl}  
\usepackage{array}
\usepackage{multirow}
\usepackage{bm} 
\newtheorem{assumption}{\textbf{Assumption}}
\newtheorem{lem}{\textbf{Lemma}}
\newtheorem{theorem}{\textbf{Theorem}}
\newtheorem{remark}{\textbf{Remark}}

\title{\LARGE \bf
Incremental Model Predictive Control for Reference Tracking of Nonlinear Anesthesia Dynamics$^{\star}$
}

\author{Maxim Raymond$^{1}$, Kaouther Moussa$^{1,2}$, Teodoro Alamo$^3$, Mirko Fiacchini$^4$, Jimmy Lauber$^{1,2}$ % 
\thanks{$^{\star}$This work was supported by the Clinical project, funded by the ANR (French National Research Agency) under grant ANR-24-CE45-4255.}% <-this % stops a space
\thanks{$^{1}$UPHF, CNRS, UMR 8201 - LAMIH, F-59313 Valenciennes, France, {\tt \small \{maxim.raymond,kaouther.moussa, jimmy.lauber\}@uphf.fr}}%
\thanks{$^2$ INSA Hauts-de-France, F-59313, Valenciennes, France.}
\thanks{$^3$ Department of Systems Engineering and Automation, Universidad de Sevilla, Sevilla, Spain, {\tt \small talamo@us.es}}
\thanks{$^4$ Univ. Grenoble Alpes, CNRS, Grenoble INP, GIPSA-lab, 38000 Grenoble, France, {\tt \small mirko.fiacchini@gipsa-lab.fr}}
}%

\begin{document}

\maketitle
\thispagestyle{empty}
\pagestyle{empty}

%%%%%%%%%%%%%%%%%%%%%%%%%%%%%%%%%%%%%%%%%%%%%%%%%%%%%%%%%%%%%%%%%%%%%%%%%%%%%%%%
\begin{abstract}
Anesthesia dynamics are commonly described by Wiener models, consisting of linear dynamics followed by a static nonlinearity. This nonlinear output complicates the derivation of theoretical guarantees on closed-loop stability and convergence. Model Predictive Control (MPC) is particularly suitable for anesthesia control, as it explicitly accounts for constraints in the control design. In this paper, we propose an incremental MPC formulation for tracking a desired reference. The proposed approach exploits the structure of the nonlinear output equation to reformulate the control problem within a linear framework. An incremental model is then employed to avoid the explicit computation of steady states and inputs, which may be inaccurate in the presence of model uncertainties. The proposed controller is evaluated through simulations involving 1000 virtual patients and compared with a benchmark MPC formulation. Results show improved steady-state reference tracking while maintaining satisfactory disturbance rejection.
\end{abstract}

%%%%%%%%%%%%%%%%%%%%%%%%%%%%%%%%%%%%%%%%%%%%%%%%%%%%%%%%%%%%%%%%%%%%%%%%%%%%%%%%
\section{Introduction}
Several mathematical models of anesthesia have been developed using clinical data, including those proposed in \cite{eleveldAllometricModelRemifentanil2017}, \cite{mintoInfluenceAgeGender1997}, and \cite{schniderInfluenceAgePropofol1999}. These models typically consist of two components: a compartmental pharmacokinetic (PK) model describing drug distribution and elimination in the body, followed by a pharmacodynamic (PD) model relating drug concentrations to their effects on the patient. The latter is commonly represented by a nonlinear Hill function.

Model Predictive Control (MPC) has been widely investigated for anesthesia control, primarily due to its ability to explicitly handle constraints. However, the nonlinear nature of anesthesia dynamics makes it challenging to establish theoretical guarantees on closed-loop stability and convergence. In \cite{ionescuNonlinearDynamicsPatients2015}, the authors proposed a linear approximation of the output dynamics, whose validity is restricted to a neighborhood of the operating point. Alternatively, nonlinear MPC approaches can be employed to directly account for the nonlinear dynamics, as in \cite{aubouin-pairaultComparisonMultipleKalman2024}. Nevertheless, such approaches do not necessarily provide theoretical guarantees on closed-loop stability and convergence.

As anesthesia dynamics are described by physiological models, they are inherently subject to significant parametric uncertainties. To address this issue, the authors in \cite{queinnecSwitchedControlStrategy2019} proposed a robust control strategy considering   only propofol as an anesthetic agent. However, the proposed approach requires the computation of steady states and inputs, which can be sensitive to model mismatch. In \cite{raymondMPCTrackingAnesthesia2025}, we proposed an MPC-based tracking approach relying on an affine reformulation of the nonlinear Hill function through its inversion. Although this formulation facilitates the control of nonlinear anesthesia dynamics, theoretical guarantees on closed-loop convergence to the desired reference were not established in that work.

The anesthesia model that we consider in this paper involves the administration of two intravenous drugs: propofol, which induces hypnosis, and remifentanil, which is used for its analgesic effect. The depth of hypnosis is assessed using the Bispectral Index (BIS), an indicator derived from electroencephalographic (EEG) signals that ranges from 0 (suppression of cerebral electrical activity) to 100 (fully awake).

Building on the affine reformulation introduced in \cite{raymondMPCTrackingAnesthesia2025}, this work proposes a control approach that ensures asymptotic convergence of the original nonlinear system to the desired reference, while relying on a simplified linear model for controller design. The incremental formulation avoids the explicit computation of steady states and input, while providing robustness to uncertainties in certain pharmacodynamic parameters. Furthermore, the proposed formulation does not require imposing a fixed ratio between the two drug infusion rates, thus retaining the additional degrees of freedom offered by the multi-input system. Finally, the integration of state estimation aims to mitigate the effects of model uncertainties, with the resulting robustness evaluated through numerical simulations.

The paper is organized as follows, in Section~\ref{section:model} both the classical and incremental formulation of the anesthesia model are presented followed by the problem statement.
The incremental MPC formulation is described in Section~\ref{section:control}. 
Section~\ref{section:results} details the simulation setup and illustrates the results. Finally, perspectives and future research directions are discussed in Section~\ref{section:conclusion}.   

%%%%%%%%%%%%%%%%%%%%%%%%%%%%%%%%%%%%%%%%%%%%%%%%%%%%%%%%%%%%%%%%%%%%%%%%%%%%%%%%
\section{Dynamical Model and Problem Formulation}\label{section:model}
The anesthesia process is described by a physiological model consisting of two components: a pharmacokinetic model represented by a compartmental system, followed by a pharmacodynamic model characterized by a nonlinear output function.

\subsection{Compartmental System}
The compartmental system is separated into 4 parts. The first three correspond to the central (blood) and the two peripheral compartments (fat, muscles).
The fourth is a virtual compartment representing the effect-site of the drug, it can be considered as the brain for the case of anesthesia.
This system allows describing the drug circulation in the body.

In \cite{zabiNewApproachControl2015}, it has been mentioned that the response time of the fat and muscles compartments can be 10 to 100 times slower compared to the blood and effect-site.
Therefore, the anesthesia model corresponds to a singularly perturbed system with a multi-time scale dynamics. 
As a consequence, the model can be divided into fast compartments (blood and effect-site) and  slow ones (fat and muscles).
While control design often aims to accelerate the slowest system dynamics, the output in anesthesia depends directly on the effect-site drug concentrations, which are associated with the faster dynamics. Therefore, the control objective is primarily to accelerate the response of the fast subsystem rather than that of the slow peripheral compartments.

As a consequence, in \cite{zabiNewApproachControl2015}, the authors proposed to consider the effect of the slow states on the fast ones as a slowly varying disturbance, 
which allows to neglect the slow dynamics.
To model this, we introduce in the dynamics an additive disturbance on the state denoted $d(t) \in \mathbb{R}$. 

\begin{align*}
    \begin{pmatrix}
        \dot{x}_{1}(t) \\
        \dot{x}_{2}(t)
    \end{pmatrix}
    & =
    \begin{pmatrix}
        -(k_{10}+k_{12}+k_{13}) & 0 \\
        k_{e}  & -k_{e}
    \end{pmatrix}
    \begin{pmatrix}
        x_{1}(t) \\
        x_{2}(t)
    \end{pmatrix} \\
    & +
    \begin{pmatrix}
        \frac{1}{V_{1}} \\
        0
    \end{pmatrix}
    u(t)
    + F d(t),
\end{align*}
with $x_1, x_2$ being respectively the drug concentration in the blood and effect-site, and $F$ the disturbance matrix.
Each drug has its own compartmental model that is independent. 
For propofol and remifentanil, the model parameters are taken respectively from \cite{eleveldPharmacokineticPharmacodynamicModel2018} and \cite{eleveldAllometricModelRemifentanil2017}.
The combined model of both drugs circulation dynamics is given by:
\begin{equation*}
    \dot{x}(t) = A x(t) + B u(t) + F d(t)
\end{equation*}
with,
\begin{equation*}
    A = \begin{pmatrix}
        A_p & 0_{2 \times 2} \\
        0_{2 \times 2} & A_r
        \end{pmatrix}\! ,
    B = \begin{pmatrix}
        B_p & 0_{2 \times 1} \\
        0_{2 \times 1} & B_r
    \end{pmatrix}\! ,
    x(t) = \begin{pmatrix} x_1^p(t) \\ x_2^p(t) \\ x_1^r(t) \\ x_2^r(t)\end{pmatrix}\! , 
\end{equation*}
where $A_p, A_r, B_p, B_r$ are the state and input matrices corresponding to propofol and remifentanil related dynamics,
$x^p$ and $x^r$ the fast states of propofol and remifentanil 
and $u(t)=(u_p(t),u_r(t))^T$ is the infusion rate of both drugs.

\subsection{Nonlinear output map}
The compartmental system is followed by the interaction model which maps the drugs effect-site concentrations to their corresponding Bispectral Index (BIS) value.
The interaction model considered here is the one from \cite{bouillonPharmacodynamicInteractionPropofol2004}, consisting of the nonlinear Hill function:
\begin{equation*}
BIS(t) = E_0 \left(1 - \dfrac{\left(U_{p}(t)+U_{r}(t)\right)^\gamma}{1+\left(U_{p}(t)+U_{r}(t)\right)^\gamma}\right), 
\end{equation*}
\begin{equation*}
    \begin{matrix}		
    U_{p}(t) = \dfrac{x^p_2(t)}{C_{e_{50}}^p}, \quad U_{r}(t) = \dfrac{x^r_2(t)}{C_{e_{50}}^r},
    \end{matrix}
\end{equation*}
with $E_0$ corresponding to the parameter denoted as \textit{Baseline} in \cite{bouillonPharmacodynamicInteractionPropofol2004},
$C_{e_{50}}^p$ and $C_{e_{50}}^r$ being the half-effect concentration of propofol and remifentanil, and $\gamma$ the slope of the Hill function.

As in \cite{raymondMPCTrackingAnesthesia2025}, the nonlinear Hill function can be reformulated to express the tracking of a constant BIS reference $r_{BIS}$ in terms of a transformed output that is linear in the effect-site concentrations:
\begin{equation}
    y(t) = \dfrac{x^p_2(t)}{C_{e50}^p} + \dfrac{x^r_2(t)}{C_{e50}^r},
    \label{Eq:output_func}
\end{equation}
where $y(t)$ denotes the transformed output. The corresponding reference is then given by
\begin{equation}\label{Eq:Linear_ref}
    r = \left( \dfrac{E_0 - r_{BIS}}{r_{BIS}} \right)^{\frac{1}{\gamma}}.
\end{equation}

\subsection{Discretized Anesthesia model}
By combining the obtained compartmental model and nonlinear output, and discretizing the whole system using the forward Euler method,

the resulting model can be expressed as:
\begin{equation}
    \begin{aligned}
    x_{k+1} & = A_d x_k + B_d u_k + F_d d_k, \\
    y_k & = C_d x_k,
    \end{aligned}
    \label{model:nominal}
\end{equation}
where
$C_d = \begin{pmatrix} 0 & \dfrac{1}{C_{e50}^p} & 0 & \dfrac{1}{C_{e50}^r} \end{pmatrix}$,
$y_k$ is the discrete-time counterpart of the transformed output $y(t)$ defined in \eqref{Eq:output_func}, and $A_d$, $B_d$, and $F_d$ are the discretized system matrices.

\begin{assumption}\label{ass:disturbance}
  The disturbance $d_k$ is assumed to be constant, i.e., $\Delta d_k = 0$.
\end{assumption}

Although the slow compartment dynamics induce variations in $d_k$, their relatively slow evolution justifies Assumption~\ref{ass:disturbance}.
Under this assumption, the incremental formulation presented in the following subsection eliminates the disturbance term from the system dynamics.
Furthermore, this formulation allows the control inputs to be adjusted incrementally, consistently with the clinical practice of anesthesiologists.

%%%%%%%%%%%%%%%%%%%%%%%%%%%%%%%%%%%%%%%%%%%%%%%%%%%%%%%%%%%%%%%%%%%%%%%%%%%%%%%%
\subsection{Problem formulation}

In clinical practice, anesthesiologists regulate anesthesia by adjusting drug infusion rates. To reflect this practice, we adopt an incremental formulation of the anesthesia model.
The incremental system can be defined using $\Delta x_k$, $\Delta u_k$ and $\Delta d_k$ which are the backward difference of $x_k$, $u_k$ and $d_k$, with respect to time. The incremental dynamics are then given by:
\begin{equation*}
    \Delta x_{k+1} = A_d \Delta x_k + B_d \Delta u_k + F_d \Delta d_k.
\end{equation*}

Under Assumption~\ref{ass:disturbance}, the following simplified system is obtained:
\begin{equation}
    \begin{aligned}
        \Delta x_{k+1} & = A_d \Delta x_k + B_d \Delta u_k \\
        \Delta y_k & = C_d \Delta x_k.
    \end{aligned}
    \label{Eq:incremental_model}
\end{equation}

%%%%%%%%%%%%%%%%%%%%%%%%%%%%%%%%%%%%%%%%%%%%%%%%%%%%%%%%%%%%%%%%%%%%%%%%%%%%%%%%

Consider the LTI system \eqref{Eq:incremental_model} in a reference-tracking setting, subject to hard input constraints $u_k \in \mathcal{U}$, for all $k \in \mathbb{N}$.
Conventional model-based tracking approaches typically require the explicit computation of steady-state values for both states and inputs.
However, due to the physiological nature of anesthesia dynamics, the system matrices are subject to significant uncertainties, potentially leading to inaccurate equilibrium computations and, consequently, steady-state tracking errors.

Let $r_{BIS}$ denote the desired BIS reference and $r$ its corresponding transformed reference, defined in \eqref{Eq:Linear_ref}.
The objective of this paper is to design an incremental controller for the original nonlinear anesthesia system that achieves reference tracking despite uncertainties in the pharmacodynamic parameters and external disturbances, while ensuring satisfaction of the input constraints.

%%%%%%%%%%%%%%%%%%%%%%%%%%%%%%%%%%%%%%%%%%%%%%%%%%%%%%%%%%%%%%%%%%%%%%%%%%%%%%%%
\section{Incremental MPC}\label{section:control}
During the surgery, in order to keep the patient BIS inside the safe zone $[40, 60]$, the anesthesiologist adapts the drug rates incrementally.
The objective of the induction phase is to reach the desired zone while avoiding a large undershoot.
Once the patient BIS reaches the targeted interval, the maintenance phase starts, where the objective 
is to maintain the patient BIS within the safe range despite disturbances induced by surgical manipulations.
To improve disturbance rejection, the BIS reference is set to the midpoint of the safe range, i.e., $r_{BIS} = 50$.

To mitigate the large BIS undershoot commonly observed during the induction phase, we introduce a first-order filter on the reference $r$, rather than tracking it directly:
\begin{equation*}
    r_k = \alpha_r r_{k-1} + (1 - \alpha_r) r.
\end{equation*}
Since anesthesia induction is typically expected to take no more than 5 minutes, $\alpha_r$ is selected such that $r_k$ approaches $r$ within a comparable time frame.
The proposed control strategy relies on the definition of a tracking error model, which enables the incorporation of integral action into the controller.

%%%%%%%%%%%%%%%%%%%%%%%%%%%%%%%%%%%%%%%%%%%%%%%%%%%%%%%%%%%%%%%%%%%%%%%%%%%%%%%%
\subsection{Error model}\label{subsection:error_model}
We first characterize the system at equilibrium:
\begin{equation}
    \begin{aligned}
    x_e & = A_d x_e + B_d u_e + F_d d_e, \\
    r & = C_d x_e.
    \end{aligned}
    \label{Eq:equilibrium_dyn}
\end{equation}
Using \eqref{Eq:equilibrium_dyn}, the error model can be derived:
\begin{align*}
    x_{k+1} - x_e & = A_d (x_k - x_e) + B_d (u_k - u_e) + F_d (d_k - d_e), \\
    r_k - r & = \alpha_r (r_{k-1} - r), \\
    e_k & = y_k - r = C_d (x_k - x_e),
\end{align*}
with $e_k$ being the error between the output and $r$.
As previously mentioned, we aim to track the smoothed reference $r_k$. Therefore, we define the following error:
\begin{equation}\label{Eq:error_k}
    e^r_k = C_d x_k - r_k = e_k - (r_k - r),
\end{equation}
standing for the error between the output and $r_k$.

To reflect the incremental adjustments made by anesthesiologists in clinical practice, we formulate the control problem using the incremental error dynamics.
Since the equilibrium values $x_e$, $u_e$, and $d_e$ are constant, their backward differences vanish.
Consequently, $\Delta(x_k-x_e) = \Delta x_k$, $\Delta(u_k-u_e) = \Delta u_k$, and $\Delta(d_k-d_e) = \Delta d_k$.
The backward differences of the reference deviation $r_k-r$ and the tracking error $e_k^r$ are then given by:
\begin{align*}
    \Delta(r_k-r) &= (r_k-r)-(r_{k-1}-r) = \Delta r_k,\\
    \Delta e_k^r &= (y_k-r_k)-(y_{k-1}-r_{k-1}),\\
    &= C_d\Delta x_k-\Delta r_k.
\end{align*}
with the dynamics of $\Delta r_k$ being $\Delta r_{k+1} = \alpha_r \Delta r_k$.

Since stabilizing $\Delta e_k^r$ alone may lead to a nonzero steady-state tracking error, we instead aim to stabilize $e_k^r$.
Let us define the dynamics of $e^r_k$ based on the incremental system~\eqref{Eq:incremental_model}:
\begin{align*}
    e^r_{k+1} & = \Delta e^r_{k+1} + e^r_k, \\
    e^r_{k+1} & = C_d A_d \Delta x_k + CB \Delta u_k - \alpha \Delta r_k + e^r_k.
\end{align*}

The resulting system dynamics used for controller synthesis are given by:
\begin{align*}
    \Delta x_{k+1} & = A_d \Delta x_k + B_d \Delta u_k, \\
    \Delta r_{k+1} & = \alpha_r \Delta r_k, \\
    e^r_{k+1} & = C_d A_d \Delta x_k + C_d B_d \Delta u_k - \alpha \Delta r_k + e^r_k.
\end{align*}

An important advantage of this formulation is that it avoids the explicit computation of the system equilibrium point,
which may be highly sensitive to model uncertainties and external disturbances. Therefore, this feature enhances the robustness
and practical applicability of the proposed approach.

Although the proposed controller is designed to track the transformed output
$y_k$, the ultimate objective is to ensure convergence of the original
nonlinear $BIS_k$ output to its desired reference $r_{BIS}$.
Moreover, the transformed reference $r$ depends on the pharmacodynamic
parameters, which are subject to uncertainties.
It is therefore important to establish the relationship between the
convergence of the transformed tracking error and that of the
original $BIS_k$ output. 

We first introduce the following functional
assumption on the estimation process.

\begin{assumption}[Asymptotic output reconstruction]\label{Ass:output_reconstruction}
Let $\hat x_k$ denote the estimated state and $\hat y_k=C_d\hat x_k$
the corresponding estimated transformed output. The estimator is assumed
to satisfy
\begin{equation}
    \lim_{k\to\infty}
    \left[
        \hat y_k-
        \left(\frac{E_0-BIS_k}{BIS_k}\right)^{\!1/\gamma}
    \right]=0,
    \label{eq:asymptotic_output_reconstruction}
\end{equation}
where $E_0$ and $\gamma$ are the parameters used to compute $r$.
\end{assumption}

The convergence result can then be stated in the following lemma.
\begin{lem}\label{Lemma:error}
Let $\gamma>0$, $0<r_{BIS}<E_0$, and
$0<BIS_k\leq E_0$ for all $k \in \mathbb N$. Under Assumption~\ref{Ass:output_reconstruction},
consider the reference at its constant value $r$ and define the estimated
tracking error $\hat e_k^r=\hat y_k-r$. Then,
\begin{equation}
    \lim_{k\to\infty}\hat e_k^r=0
    \quad\Longleftrightarrow\quad
    \lim_{k\to\infty}BIS_k=r_{BIS}.
\end{equation}
\end{lem}

\begin{proof}
Let $\phi(s)=\bigl((E_0-s)/s\bigr)^{1/\gamma}$, so that
$r=\phi(r_{BIS})$. We have
\begin{equation}
    \hat e_k^r
    =\underbrace{\hat y_k-\phi(BIS_k)}_{\longrightarrow 0}
      +\phi(BIS_k)-\phi(r_{BIS}).
\end{equation}
The first term vanishes by Assumption~\ref{Ass:output_reconstruction}. Since the inverse Hill transformation is continuous and strictly
monotonic on $(0,E_0]$, convergence of the transformed output to
$\phi(r_{\mathrm{BIS}})$ is equivalent to convergence of  $BIS_k$
to $r_{\mathrm{BIS}}$.

\end{proof}

\begin{remark}
Assumption~\ref{Ass:output_reconstruction} requires asymptotic reconstruction of the
transformed output, and not exact state estimation. In particular, the state
estimation error may converge to a nonzero constant provided that this
bias does not affect the reconstructed output. Convergence of the
estimated state increments alone does not guarantee this property.
\end{remark}

As a result of Lemma~\ref{Lemma:error}, the convergence of the transformed tracking error to zero is equivalent to the asymptotic convergence of $BIS_k$ 
in the original nonlinear system to its desired reference. 
Importantly, this equivalence does not require the exact knowledge of $\gamma$ and $E_0$, provided that the same transformation is applied to both
the measured BIS and its reference.

We then define the augmented state vector used for controller synthesis as follows:
\begin{equation*}
    z_k = \begin{bmatrix} \Delta x_k^T & \Delta r_k & e^r_k \end{bmatrix}^T,
\end{equation*} 
with the corresponding dynamics:
\begin{equation}\label{Eq:z_dyn}
    \begin{aligned}
    z_{k+1} & = \begin{bmatrix} A_d & 0 & 0 \\ 0 & \alpha_r I & 0 \\ C_d A_d & -\alpha_r I & I \end{bmatrix} z_k + \begin{bmatrix} B_d \\ 0 \\ C_d B_d \end{bmatrix} \Delta u_k, \\
    e^r_k & = \begin{bmatrix} 0 & 0 & I \end{bmatrix} z_k.
    \end{aligned}
\end{equation}

%%%%%%%%%%%%%%%%%%%%%%%%%%%%%%%%%%%%%%%%%%%%%%%%%%%%%%%%%%%%%%%%%%%%%%%%%%%%%%%%
\subsection{MPC formulation}\label{subsection:incremental_mpc}
MPC techniques are commonly used for the control of anesthesia because of their ability to handle constraints by design.
%The critical constraints for anesthesia are the maximum allowable infusion rates of propofol and remifentanil.
The hard constraints in anesthesia include the maximum allowable infusion rates of propofol and remifentanil.
Since we use an incremental model, defining constraints on the input $\Delta u_k$ will not guarantee the satisfaction of the constraints on $u_k$.
% To overcome this issue, we extend the state $z_k$ with the input $u_{k-1}$ as follows: 
To overcome this issue, the terminal positive invariant set $\mathcal{X}$ and the constraints of the MPC are considered on
the extended state $\xi_k$, including the control input at time $k-1$, that is defined as follows: 
\begin{equation*}
    \xi_k = \begin{bmatrix} \Delta x_k^T & \Delta r_k & e^r_k & u_{k-1}^T \end{bmatrix}^T,
\end{equation*}
and the resulting dynamics are:
\begin{equation}\label{Eq:extended_system}
    \begin{aligned}
        \xi_{k+1} & = \begin{bmatrix} A_d & 0 & 0 & 0 \\ 0 & \alpha_r I & 0 & 0 \\ C_d A_d & -\alpha_r I & I & 0 \\ 0 & 0 & 0 & I \end{bmatrix} \xi_k + \begin{bmatrix} B_d \\ 0 \\ C_d B_d \\ I \end{bmatrix} \Delta u_k, \\
        e^r_k & = \begin{bmatrix} 0 & 0 & I & 0 \end{bmatrix} \xi_k.
    \end{aligned}
\end{equation}

Using this extended system, we can then define the optimization problem as follows:
\begin{equation}\label{Eq:opt_mpc}
    \begin{aligned}
        \underset{\Delta u}{\text{min}} & \sum_{k=0}^{N-1} \,\, \lVert \xi_k \rVert_{Q_{ct}}^2 \, + \, \lVert \Delta u_k \rVert_{R_{ct}}^2 \, + \, \lVert \xi_N \rVert_{P_{ct}}^2 \\
        \text{s.t.} \: & \xi_0 = \xi(t), \\
        & \xi_{k+1} = A_{\xi} \xi_k + B_{\xi} \Delta u_k, \quad k \in \mathbb{N}_{[0, N-1]}, \\
        & (\xi_k, \Delta u_k) \in \mathcal{Z}, \qquad \qquad \,\,\, k \in \mathbb{N}_{[0, N-1]},\\
        & \xi_N \in \mathcal{X},
    \end{aligned}
\end{equation}
where $\Delta u$ denotes the sequence of control increments over the prediction horizon, $Q_{ct} \succeq 0$, $R_{ct} \succ 0$, and $P_{ct} \succeq 0$ are the weighting matrices, $\mathcal{Z}$ represents the state and input constraint set, $\mathcal{X}$ is an invariant terminal set, and $A_\xi$ and $B_\xi$ are the state and input matrices of the extended system \eqref{Eq:extended_system}, respectively. Then, the MPC controller is obtained by recursively solving the optimization problem \eqref{Eq:opt_mpc} in a receding horizon fashion.
%is obtained by applying a receding horizon control to \eqref{Eq:opt_mpc}.

For the theoretical analysis, we consider a constant reference,
$r_k=r$, and thus $\Delta r_k=0$. The general formulation
\eqref{Eq:extended_system}--\eqref{Eq:opt_mpc} is retained for the
numerical implementation, including the use of a filtered reference.
In the constant-reference case, we introduce the reduced variables
\begin{equation*}\label{eq:reduced_state}
    \bar z_k = \begin{bmatrix}\Delta x_k\\ e_k^r\end{bmatrix},
    \qquad
    \bar\xi_k = \begin{bmatrix}\bar z_k\\u_{k-1}\end{bmatrix}.
\end{equation*}
Their dynamics are given by
\begin{equation*}
\bar z_{k+1} = \bar A_z\bar z_k+\bar B_z\Delta u_k,
\end{equation*}
\begin{equation}
    \bar\xi_{k+1} = \bar A_\xi\bar\xi_k+\bar B_\xi\Delta u_k,\label{eq:reduced_dynamics}
\end{equation}
where
\begin{equation*}
    \begin{gathered}
    \bar A_z=\begin{bmatrix}A_d&0\\C_dA_d&1\end{bmatrix},\qquad
    \bar B_z=\begin{bmatrix}B_d\\C_dB_d\end{bmatrix},\\
    \bar A_\xi=\begin{bmatrix}\bar A_z&0\\0&I_2\end{bmatrix},\qquad
    \bar B_\xi=\begin{bmatrix}\bar B_z\\I_2\end{bmatrix}.
    \end{gathered}
\end{equation*}
The set of equilibria of \eqref{eq:reduced_dynamics} under zero input
increment is
\begin{equation*}\label{eq:attractor_A}
    \mathcal A:=\left\{
    \bar\xi=\begin{bmatrix}\bar z\\u^-\end{bmatrix}:
    \bar z=0,\quad u^-\in\mathcal U\right\},
\end{equation*}
where $\mathcal U \subset \mathbb R^2$ denotes the control constraints set. We denote the distance to $\mathcal A$ by
\begin{equation*}\label{eq:distance_A}
    \|\bar\xi\|_{\mathcal A}:=
    \min_{\eta\in\mathcal A}\|\bar\xi-\eta\|.
\end{equation*}

For every admissible $\bar\xi$ with $u^-\in\mathcal U$, it follows
that $\|\bar\xi\|_{\mathcal A}=\|\bar z\|$. In particular, for a
positive definite matrix $Q$, we use the weighted distance
\begin{equation*}\label{eq:weighted_distance_A}
    \begin{aligned}
    \|\bar\xi\|_{Q,\mathcal A}^2
    &:=\min_{\eta\in\mathcal A}
    \|\bar\xi-\eta\|_{\operatorname{diag}(Q,I_2)}^2\\
    &=\|\bar z\|_Q^2,\qquad u^-\in\mathcal U.
    \end{aligned}
\end{equation*}
Using the reduced
variables defined above, the MPC optimization problem
becomes
\begin{equation}
\label{Eq:reduced_mpc}
\begin{aligned}
\min_{\Delta u} \quad &
\sum_{k=0}^{N-1}
\left(
\|\bar{\xi}_k\|_{Q,\mathcal A}^2
+
\|\Delta u_k\|_R^2
\right)
+
\|\bar{\xi}_N\|_{P,\mathcal A}^2\\
\text{s.t.}\quad
&\bar{\xi}_0=\bar{\xi}(t),\\
&\bar{\xi}_{k+1}
=\bar{A}_{\xi}\bar{\xi}_k
+\bar{B}_{\xi}\Delta u_k, \quad k \in \mathbb{N}_{[0, N-1]}\\
&(\bar{\xi}_k,\Delta u_k)\in\bar{\mathcal Z},
\qquad \quad \: \: \: \: \: \: \; k \in \mathbb{N}_{[0, N-1]} \\
&\bar{\xi}_N\in\bar{\mathcal X}_f.
\end{aligned}
\end{equation}
where $Q, P \succ0$ denote the weighting matrices on $\bar z$, and $R\succ0$
is the input-increment weighting matrix. The dynamics and constraints
of this reduced MPC problem are obtained by restricting
\eqref{Eq:opt_mpc} to \eqref{eq:reduced_dynamics}.

We next establish a terminal invariance property for the
constant reference system.
\begin{lem}
\label{lem:terminal_invariance}
Let $\mathcal U\subset\mathbb R^2$ be compact and convex,
and let $K$ be such that
$A_K=\bar A_z+\bar B_zK$ is Schur.
Consider a compact, convex set $\mathcal Z_f$ containing
the origin and satisfying
$A_K\mathcal Z_f\subseteq\mathcal Z_f$.
Define
\begin{equation*}
H=K(I-A_K)^{-1},\qquad
\mathcal V_f=\mathcal U\ominus(-H\mathcal Z_f),
\end{equation*}
and suppose $\mathcal V_f\neq\emptyset$.
Consider the set
\begin{equation*}
\bar{\mathcal X}_f^0=
\left\{
\bar\xi_k:
\bar z_k\in\mathcal Z_f,\;
u_{k-1}+H\bar z_k\in\mathcal V_f
\right\}.
\end{equation*}
Then, the set
\begin{equation*}
\bar{\mathcal X}_f=
\operatorname{conv}
\left(\bar{\mathcal X}_f^0\cup\mathcal A\right)
\end{equation*}
is compact, convex and positively invariant under
$\Delta u_k=K\bar z_k$, while satisfying the input
constraints and $\mathcal A\subseteq\bar{\mathcal X}_f$.
\end{lem}

\begin{proof}
Introduce $v_k=u_{k-1}+H\bar z_k$ and suppose that $\bar{\xi}_k \in \bar{\mathcal{X}}_f^0$. Under the terminal
feedback $\Delta u_k=K\bar z_k$, we obtain
\begin{align}
\bar z_{k+1}&=A_K\bar z_k, \label{Eq:closedDynZ}\\
v_{k+1}
&=u_k+H\bar z_{k+1} \nonumber\\
&=u_{k-1}+K\bar z_k+HA_K\bar z_k
=v_k, \nonumber
\end{align}
where the last equality follows from
$H(I-A_K)=K$.
Consequently, for all $j\geq 0$,
\begin{equation*}
\bar z_{k+j}=A_K^j\bar z_k,\qquad
v_{k+j}=v_k.
\end{equation*}
Since $\mathcal Z_f$ is positively invariant for \eqref{Eq:closedDynZ}, then
$A_K^j\bar z_k\in\mathcal Z_f$ and $v_{k+j} \in \mathcal{V}_f$. Moreover, by the
definition of $\mathcal V_f$,
\begin{equation*}
u_{k+j-1}
=v_k-HA_K^j\bar z_k\in\mathcal U.
\end{equation*}
Thus, $\bar{\mathcal X}_f^0$ is positively invariant
and satisfies the input constraints.

Now, consider the extended closed-loop dynamics
under $\Delta u_k=K\bar z_k$:
\begin{equation*}
\bar\xi_{k+1}=F_K\bar\xi_k,\qquad
F_K=
\begin{bmatrix}
A_K & 0\\
K & I_2
\end{bmatrix}.
\end{equation*}
By construction,
$
F_K\bar{\mathcal X}_f^0
\subseteq\bar{\mathcal X}_f^0.
$

Furthermore, every point of $\mathcal A$ is an
equilibrium under the terminal feedback, yielding $
F_K\mathcal A=\mathcal A.
$

Since $F_K$ is linear, it preserves convex
combinations. Therefore,
\begin{equation*}
\begin{aligned}
F_K\bar{\mathcal X}_f
&=F_K\operatorname{conv}
\left(\bar{\mathcal X}_f^0\cup\mathcal A\right)\\
&=\operatorname{conv}
\left(F_K\bar{\mathcal X}_f^0\cup F_K\mathcal A\right)\\
&\subseteq\operatorname{conv}
\left(\bar{\mathcal X}_f^0\cup\mathcal A\right)\\
&=\bar{\mathcal X}_f.
\end{aligned}
\end{equation*}
Hence, $\bar{\mathcal X}_f$ is positively invariant.

Input admissibility follows from the linearity of
$u_k=u_{k-1}+K\bar z_k$ and the convexity of
$\mathcal U$, since the input constraints hold
on both $\bar{\mathcal X}_f^0$ and $\mathcal A$.

Finally, compactness and convexity follow from
the construction of $\bar{\mathcal X}_f$ as the
convex hull of two compact sets, while
$\mathcal A\subseteq\bar{\mathcal X}_f$
holds by definition.
\end{proof}

Having established the existence of a terminal control
invariant set containing the attractor $\mathcal A$,
we now investigate the closed-loop properties of the
proposed incremental MPC. The following theorem
establishes recursive feasibility and asymptotic
stability with respect to $\mathcal A$ for the nominal
system under a constant reference. Furthermore,
convergence of the original nonlinear BIS output
is established under the assumptions of
Lemma~\ref{Lemma:error}.

\begin{theorem}
\label{th:stability}
Consider the reduced system~\eqref{eq:reduced_dynamics}
and the MPC problem~\eqref{Eq:reduced_mpc}, with
$Q\succ0$, $R\succ0$, and the terminal set
$\bar{\mathcal X}_f$ defined in
Lemma~\ref{lem:terminal_invariance}.

Assume that $\mathcal U$, $\mathcal Z_f$, and
$\bar{\mathcal Z}$ are polyhedral, with
$\mathcal U$ and $\mathcal Z_f$ compact.
Let the terminal control law $\Delta u=K\bar z$
be admissible on $\bar{\mathcal X}_f$, and
let $P\succ0$ satisfy
\begin{equation}
A_K^T P A_K-P
\preceq -(Q+K^T RK).
\label{eq:terminal_decrease}
\end{equation}
Assume also that the stage constraints enforce
$u^-\in\mathcal U$ and
$u^-+\Delta u\in\mathcal U$.

Then, for every initial condition belonging to
the MPC feasibility region $\bar{\mathcal X}_N$,
the optimization problem~\eqref{Eq:reduced_mpc}
is recursively feasible, and the attractor
$\mathcal A$ is asymptotically stable for the
resulting nominal closed-loop system, relative
to $\bar{\mathcal X}_N$. Furthermore, under Assumption~\ref{Ass:output_reconstruction} and the conditions
of Lemma~\ref{Lemma:error}, the original nonlinear
BIS output converges asymptotically to the desired
reference $r_{BIS}$.
\end{theorem}

\begin{proof}
We establish the result by verifying the conditions of Theorem~1 in
\cite{ajamiModelPredictiveControl2026}, 
which particularized the standard MPC stability
 to an
equilibrium set, based on Theorem~2.19 from \cite{bookMayneRawlings}.

\emph{Condition~1:}
The reduced dynamics are linear, and the stage
and terminal costs,
\begin{equation*}
\begin{aligned}
\ell(\bar\xi,\Delta u)
&=\bar z^T Q\bar z+\Delta u^T R\Delta u,\\
V_f(\bar\xi)&=\bar z^T P\bar z,
\end{aligned}
\end{equation*}
are continuous and nonnegative.
Moreover, every $\eta\in\mathcal A$ is an
equilibrium under $\Delta u=0$, with
$\ell(\eta,0)=V_f(\eta)=0$.
Thus, Condition~1 holds.

\emph{Condition~2:}
By assumption, the constraint set
$\bar{\mathcal Z}$ is closed.
From Lemma~\ref{lem:terminal_invariance},
$\bar{\mathcal X}_f$ is compact and contains
$\mathcal A$. Furthermore, zero input
increments are admissible on $\mathcal A$.
Hence, Condition~2 holds.

\emph{Condition~3.a:}
Lemma~\ref{lem:terminal_invariance} ensures
that $\bar{\mathcal X}_f$ is positively
invariant under the admissible terminal
control law $\Delta u=K\bar z$.
Therefore, Condition~3.a is satisfied.

\emph{Condition~3.b:}
Using the terminal cost
$V_f(\bar\xi)=\bar z^T P\bar z$
and the terminal feedback, we obtain
\begin{equation*}
\begin{aligned}
V_f(F_K\bar\xi)-V_f(\bar\xi) & =\bar z^T(A_K^T P A_K-P)\bar z\\
&\leq-\bar z^T(Q+K^T RK)\bar z\\
&=-\ell(\bar\xi,K\bar z),
\end{aligned}
\end{equation*}
where the inequality follows from
\eqref{eq:terminal_decrease}.
Thus, Condition~3.b holds.

\emph{Condition~4.a:}
Since $u^-\in\mathcal U$, the distance
to the attractor satisfies
$\|\bar\xi\|_{\mathcal A}=\|\bar z\|$.
Consequently,
\begin{equation*}
\begin{aligned}
\ell(\bar\xi,\Delta u)
&=\bar z^T Q\bar z+\Delta u^T R\Delta u\\
&\geq\lambda_{\min}(Q)\|\bar z\|^2\\
&=\lambda_{\min}(Q)
\|\bar\xi\|_{\mathcal A}^2.
\end{aligned}
\end{equation*}
Condition~4.a is therefore satisfied with
$\alpha_1(s)=\lambda_{\min}(Q)s^2$.

\emph{Condition~4.b:}
Similarly, the quadratic terminal cost satisfies
\begin{equation*}
\begin{aligned}
V_f(\bar\xi)
&=\bar z^T P\bar z\\
&\leq\lambda_{\max}(P)\|\bar z\|^2\\
&=\lambda_{\max}(P)
\|\bar\xi\|_{\mathcal A}^2.
\end{aligned}
\end{equation*}
Thus, Condition~4.b holds with
$\alpha_f(s)=\lambda_{\max}(P)s^2$.

\emph{Condition~5:}
%In this section, we denote with the superscript $*$ the optimizers $V$, $\Delta\mathbf u$ and $\bar{z}$.
Let $V_N^*$ denote the optimal value function
of~\eqref{Eq:reduced_mpc}. 
Since the dynamics are linear, the cost is quadratic,
and the constraints are polyhedral, the MPC problem
can be formulated as a strictly convex quadratic
program parameterized by the initial state
$\bar\xi$. Its optimal input sequence is therefore
continuous and piecewise affine over a finite
polyhedral partition of the convex feasibility region
$\bar{\mathcal X}_N$ \cite{BEMPORAD20023}, and hence Lipschitz
continuous.

Moreover, since
$\mathcal A\subseteq\bar{\mathcal X}_f$,
the zero-increment sequence is feasible and
optimal at every $\eta\in\mathcal A$, yielding
\begin{equation*}
V_N^*(\eta)=0,\qquad
\Delta\mathbf u^*(\eta)=0,
\end{equation*}
where $\Delta\mathbf u^*(\eta)$ denotes the optimal MPC solution at $\eta$. Let $\eta$ be
the projection of $\bar \xi$ onto $\mathcal A$.
By Lipschitz continuity, there exists
$L>0$ such that
\begin{equation*}
\begin{aligned}
\|\Delta\mathbf u^*(\bar\xi)\|
&=\|\Delta\mathbf u^*(\bar\xi)
-\Delta\mathbf u^*(\eta)\|\\
&\leq L\|\bar\xi-\eta\|\\
&=L\|\bar\xi\|_{\mathcal A}.
\end{aligned}
\end{equation*}

Since the prediction horizon is finite,
the reduced linear dynamics imply that
there exist constants $c_i>0$ such that
\begin{equation*}
\|\bar z_i^*\|
\leq c_i\|\bar\xi\|_{\mathcal A},
\qquad i=0,\ldots,N,
\end{equation*}
where $\bar{z}_i^*$ denotes the optimal predicted state at time $i$.
Using these bounds in the quadratic
objective, there exists $c>0$ satisfying
\begin{equation*}
V_N^*(\bar\xi)
\leq c\|\bar\xi\|_{\mathcal A}^2,
\qquad
\forall\bar\xi\in\bar{\mathcal X}_N.
\end{equation*}
Therefore, Condition~5 holds with
$\alpha_2(s)=cs^2$.

Since all five conditions are satisfied,
recursive feasibility and asymptotic
stability of $\mathcal A$ follow from
Theorem~2.19 in \cite{bookMayneRawlings}.

In particular, $\bar z_k\to0$, implying
$e_k^r\to0$. Under the assumptions of
Lemma~\ref{Lemma:error}, this implies
$BIS_k\to r_{BIS}$, which completes the proof.
\end{proof}

\begin{remark}
The theoretical guarantees of
Theorem~\ref{th:stability} are established
for the nominal system under a constant
reference and exact state information.
In practice, the state increments and
tracking error are reconstructed from
noisy measurements. The influence of
estimation errors on recursive feasibility
and closed-loop stability is not addressed
in this work. Similarly, robustness to
model uncertainties and time-varying
disturbances is not included in the
theoretical analysis, but is evaluated
through numerical simulations.
\end{remark}

%%%%%%%%%%%%%%%%%%%%%%%%%%%%%%%%%%%%%%%%%%%%%%%%%%%%%%%%%%%%%%%%%%%%%%%%%%%%%%%%
\section{Numerical tests}\label{section:results}

To evaluate the performance and robustness of the proposed incremental MPC, we compare it with a benchmark MPC designed using the classical anesthesia model \eqref{model:nominal}.
Both controllers are assessed on a population of virtual patients in terms of reference tracking and disturbance rejection, in the presence of significant uncertainties in the pharmacokinetic and pharmacodynamic parameters.
The disturbance $d(t)$ accounts for both the influence of the neglected slow compartment dynamics and model uncertainties.

Since the BIS is derived from electroencephalographic (EEG) signals, its measurement is subject to significant noise and artifacts, motivating the use of Kalman filtering for state estimation.
Accordingly, a Kalman filter is designed for each control formulation, estimating the state increments $\Delta x_k$ for the incremental MPC and the states $x_k$ together with the disturbance $d_k$ for the benchmark MPC.

This section is organized as follows. Section~\ref{subsection:bench_control} presents the benchmark MPC formulation, Section~\ref{subsection:simu_setup} describes the simulation setup, and Section~\ref{subsection:results} discusses the numerical results.

\subsection{Benchmark MPC}\label{subsection:bench_control}

Considering system \eqref{model:nominal}, the benchmark MPC is derived from the solution of the following optimization problem:
\begin{equation*}
    \begin{aligned}
        \underset{u}{\text{min}} & \sum_{k=0}^{N-1} \,\, \lVert C_d x_k - r \rVert_{Q_{b}}^2 \, + \, \lVert u_k - u_e \rVert_{R_{b}}^2 \, + \, \lVert C_d x_N - r \rVert_{P_{b}}^2 \\
        \text{s.t.} \: & x_0 = x(t), \, d=d(t) \\
        & x_{k+1} = A_d x_k + B_d u_k + F_d d, \quad k \in \mathbb{N}_{[0, N-1]}, \\
        & (x_k, u_k) \in \mathcal{Z}_b, \qquad \qquad \qquad \quad  \,\, k \in \mathbb{N}_{[0, N-1]},
    \end{aligned}
\end{equation*}
where $u$ is the input sequence over the horizon, $Q_{b} \succ 0, R_{b} \succ 0$ are the weighting matrices, $P_{b} \succ 0$ is the terminal weighting matrix, $\mathcal{Z}_b$ represents the states and input constraints set
and $u_e$ is the input at equilibrium, resulting from the following optimization problem:
\begin{equation*}
    \begin{aligned}
        \underset{u_e}{\text{min}} & \,\, \lVert C_d x_e - r \rVert_{Q_{e}}^2 \, + \, \lVert x_e - x_0 \rVert_{R_{e}}^2 \\
        \text{s.t.} \: & x_0 = x(t), \, d = d(t) \\
        & x_e = A_d x_e + B_d u_e + F_d d, \quad k \in \mathbb{N}_{[0, N-1]}, \\
        & (x_e, u_e) \in \mathcal{Z}_e, \qquad \qquad \qquad \, k \in \mathbb{N}_{[0, N-1]},
    \end{aligned}
\end{equation*}
where $Q_{e}, R_{e}$ are the weighting matrices and $x_e$, $u_e$ the equilibrium states and inputs.

\subsection{Simulation Setup}\label{subsection:simu_setup}
To test both controllers, the simulator developed in \cite{aubouin-pairaultPASPythonAnesthesia2023} was used as the test bench.
Patient were drawn randomly from a discrete uniform distribution with a height between $150$ and $190$ cm, a weight between $50$ and $100$ kg, 
an age between $18$ and $70$ years old, and a randomly selected gender (male or female).
For each patient, the corresponding nominal model was used in the design of the controller and estimator, while the simulator was using another system subject to uncertainties in the model parameters.
To verify the ability of the controller to reject disturbances, a step disturbance of amplitude $10$ was added between the $10^{th}$ and the $15^{th}$ minutes of the simulation. 
Finally, the output is subject to a filtered white noise.

The states were estimated every second, while we controlled every $5$ seconds with a horizon length of the MPC equal to $60$ steps.
The weighting matrices used for the incremental MPC are  $Q_{ct} = diag(\left[ 1, 1, 1, 1, 0, 10, 0, 0 \right])$ and $R_{ct} = I_2$.
For the benchmark MPC $Q_b = 100$, $R_b = 5I_2$, $Q_e = 1000$ and $R_e = I_5$  were used.
In both cases, the terminal weighting matrix was computed using the \textit{dlqr} function from the \textit{control} python library.
Tests were conducted on $1000$ different patients. 
For the comparison between both controllers, we used the constant reference $r$. 
Then, the filtered reference $r_k$ was used in the incremental MPC assessment.

\subsection{Results}\label{subsection:results}

\begin{figure}[ht]
    \centering
    \includegraphics[width=1\linewidth]{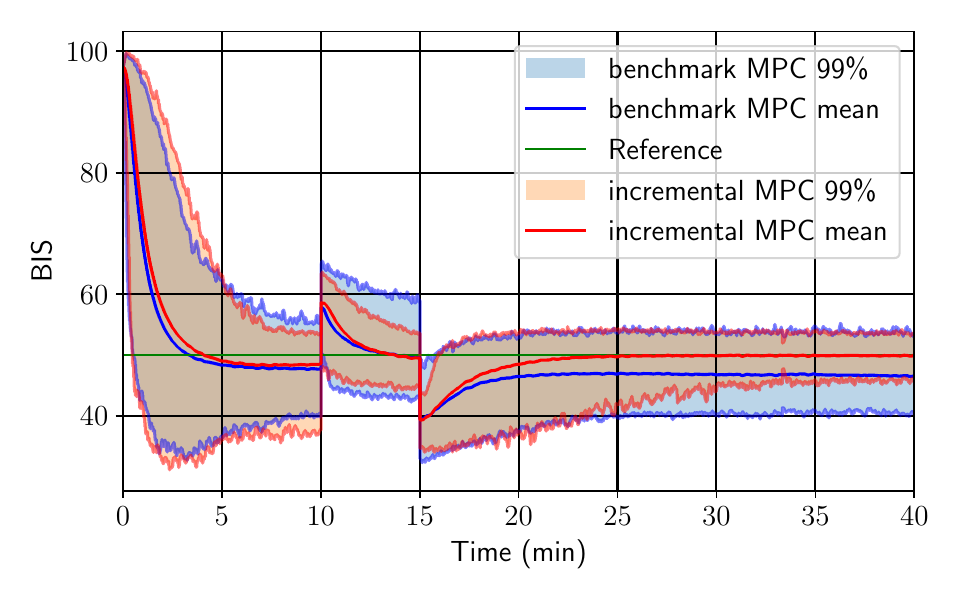}
    \caption{BIS evolution comparison on 1000 patient between the incremental and benchmark MPC}
    \label{fig:comparison_benchmark}
\end{figure}

\begin{figure}[ht]
    \centering
    \includegraphics[width=1\linewidth]{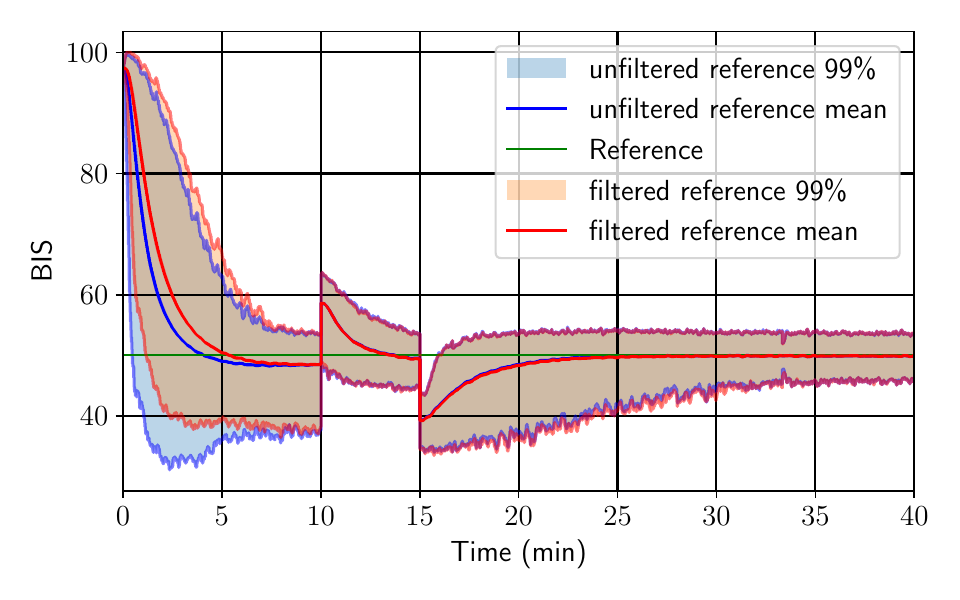}
    \caption{Comparison of BIS trajectories for 1000 virtual patients under incremental MPC with filtered ($r_k$) and unfiltered ($r$) references.
}
    \label{fig:comparison_filter}
\end{figure}

\begin{figure}[ht]
    \centering
    \includegraphics[width=1\linewidth]{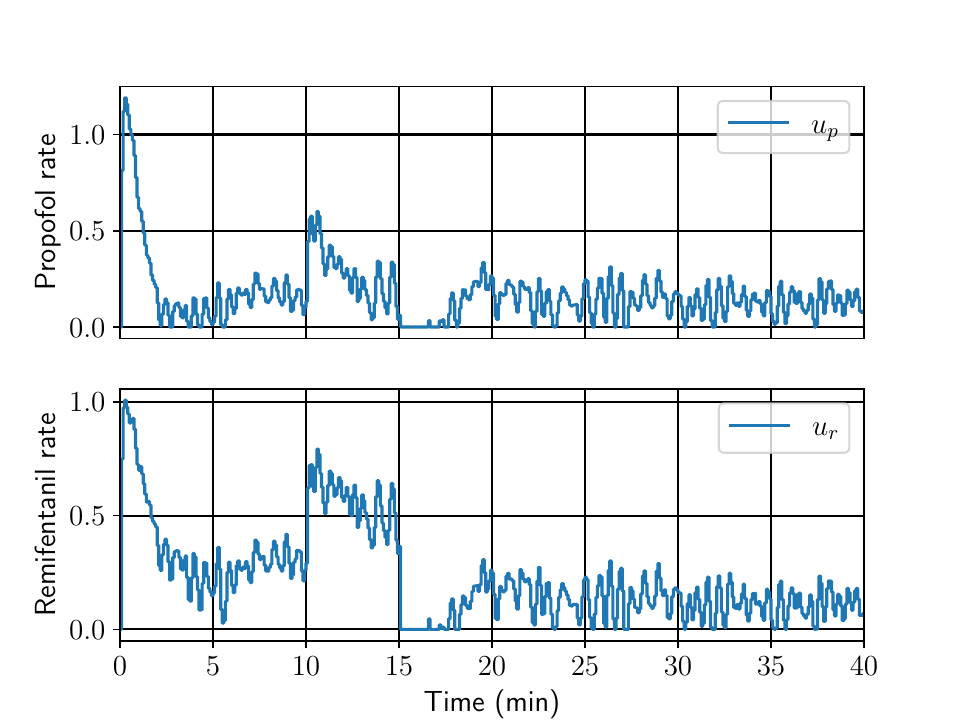}
    \caption{Representative patient control profiles}
    \label{fig:input}
\end{figure}

\begin{figure}[ht]
    \centering
    \includegraphics[width=1\linewidth]{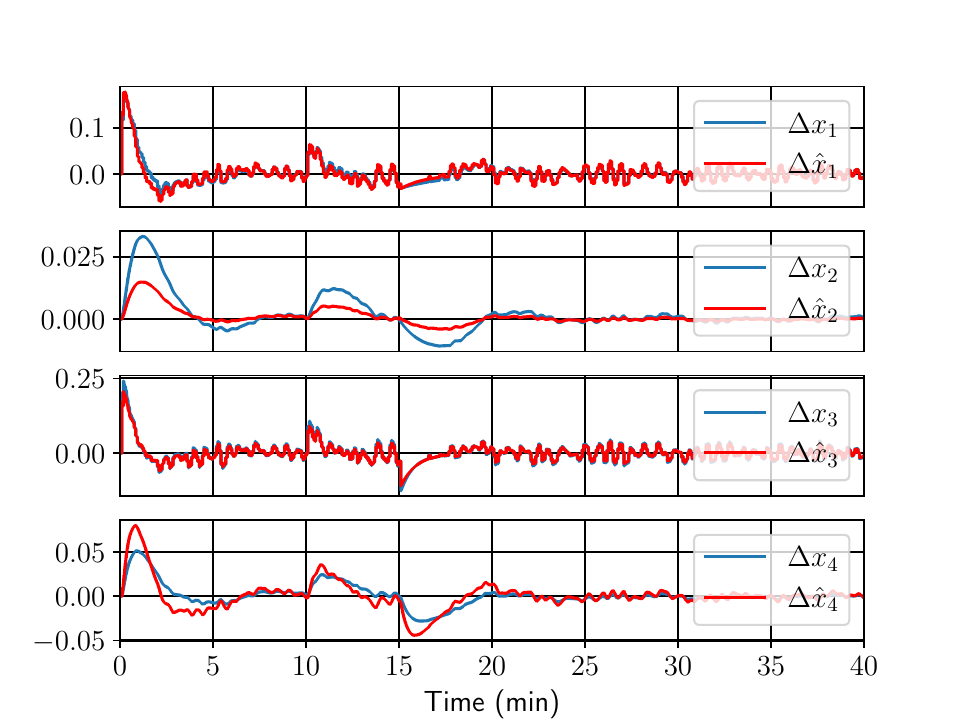}
    \caption{Representative patient state estimation}
    \label{fig:state}
\end{figure}

\begin{table}[ht]
\centering
\renewcommand{\arraystretch}{1.3}
\arrayrulecolor{black}

\resizebox{1\linewidth}{!}{
\begin{tabular}{c|rl|rl|rl}
\rowcolor{gray!35}
& \multicolumn{2}{c|}{\textbf{\textbf{IAE}}} & \multicolumn{2}{c|}{\textbf{\textbf{Lowest BIS}}} & \multicolumn{2}{c}{\textbf{\textbf{Settling time (min)}}} \\
\rowcolor{gray!35}
\multirow{-2}{*}{Controller}& mean & $\pm$ std & mean & $\pm$ std & mean & $\pm$ std  \\
\rowcolor{gray!0}
Benchmark MPC   & $11808.97$ & $\pm 3915.82$ & $37.31$ &  $\pm 3.27$  & $3.04$ & $ \pm 2.54$ \\
\rowcolor{gray!20}
Incremental MPC  &  &  & & & &\\
\rowcolor{gray!20}
(constant reference) & \multirow{-2}{*}{$9258.42$} &  \multirow{-2}{*}{$ \pm 3949.87$} & \multirow{-2}{*}{$36.89$} & \multirow{-2}{*}{$\pm 2.5$}  & \multirow{-2}{*}{$3.13$} &  \multirow{-2}{*}{$ \pm 1.77$} \\
\rowcolor{gray!0}
Incremental MPC  &  &  & & & &\\
\rowcolor{gray!0}
(filtered reference) & \multirow{-2}{*}{$10233.37$} & \multirow{-2}{*}{$ \pm 4042.41$} & \multirow{-2}{*}{$37.1$} &  \multirow{-2}{*}{$\pm 2.01$}  & \multirow{-2}{*}{$4.13$}&  \multirow{-2}{*}{$ \pm 1.68$} \\
\end{tabular}
}
\caption{Performance metrics for the three configurations. }
\label{table:performance}
\end{table}

In Fig.~\ref{fig:comparison_benchmark}, the regions labeled \textit{benchmark MPC 99\%} and \textit{incremental MPC 99\%} represent the pointwise empirical intervals containing 99\% of the simulated BIS values at each time step for the respective controllers.
The corresponding mean BIS trajectories are denoted by \textit{benchmark MPC mean} and \textit{incremental MPC mean}.
Similarly, Fig.~\ref{fig:comparison_filter} compares the performance of the incremental MPC using the filtered reference $r_k$ and the constant reference $r$, following the same notation.
The performance metrics for the three configurations are summarized in Table~\ref{table:performance}.
Finally, the control input profiles and state estimates obtained with the incremental MPC for a representative patient are illustrated in Figs.~\ref{fig:input} and~\ref{fig:state}, respectively.

During the induction phase, both controllers exhibit comparable response times, although the incremental MPC results in a slightly larger BIS undershoot.
As illustrated in Fig.~\ref{fig:comparison_filter}, this undershoot can be reduced by introducing the filtered reference $r_k$, at the expense of a longer settling time, as reported in Table~\ref{table:performance}.

During the maintenance phase, the mean BIS trajectory under incremental MPC converges to the desired reference, whereas the benchmark MPC exhibits a nonzero steady-state tracking error.
This improvement is also reflected in the lower integral absolute error (IAE) reported in Table~\ref{table:performance}.
Furthermore, Fig.~\ref{fig:comparison_benchmark} shows a smaller dispersion of BIS values across patients under incremental MPC.
Regarding disturbance rejection, both controllers achieve comparable performance in terms of the mean BIS response, while the incremental MPC maintains a narrower BIS dispersion.

%%%%%%%%%%%%%%%%%%%%%%%%%%%%%%%%%%%%%%%%%%%%%%%%%%%%%%%%%%%%%%%%%%%%%%%%%%%%%%%%
\section{Conclusion}\label{section:conclusion}

In this paper, we proposed an incremental MPC approach for anesthesia control, exploiting an equivalent linear reformulation of the nonlinear Hill output function.
The proposed formulation ensures convergence of the original nonlinear BIS output to its desired reference under the considered assumptions, including uncertainties in certain pharmacodynamic parameters.
Furthermore, the incremental model avoids the explicit computation of equilibrium states and inputs, as well as the need to estimate constant disturbances.

The proposed approach was evaluated through numerical simulations involving uncertainties in both pharmacokinetic and pharmacodynamic parameters.
The results demonstrate improved reference tracking and reduced BIS dispersion compared to a conventional MPC formulation, while maintaining satisfactory disturbance rejection.
Nevertheless, these uncertainties are not explicitly accounted for in the controller design.
Future work will therefore focus on extending the theoretical guarantees of recursive feasibility and closed-loop stability to uncertain models.
Additionally, since BIS measurement noise is not directly accounted for in the MPC formulation, the resulting control input profiles exhibit significant variations that are inconsistent with clinical practice.
Addressing this limitation constitutes another important direction for future research. 

Finally, the present work relies on an assumption regarding output reconstruction. Investigating the conditions under which this assumption remain valid, particularly in the presence of model uncertainties,  and their implications for the theoretical guarantees, constitutes an important direction for future research.

\vspace{-0.2cm}
\section*{Acknowledgment}
The authors acknowledge the use of ChatGPT (OpenAI) for language editing and text reformulation. All content was carefully reviewed and verified by the authors, who take full responsibility for the final manuscript.

%%%%%%%%%%%%%%%%%%%%%%%%%%%%%%%%%%%%%%%%%%%%%%%%%%%%%%%%%%%%%%%%%%%%%%%%%%%%%%%%

\bibliography{main_ecc}

@article{zabiNewApproachControl2015,
  title = {New Approach for the Control of Anesthesia Based on Dynamics Decoupling},
  author = {Zabi, Said and Queinnec, Isabelle and Tarbouriech, Sophie and Garcia, Germain and Mazerolles, Michel},
  year = 2015,
  journal = {IFAC-PapersOnLine},
  volume = {48},
  number = {20},
  pages = {511--516},
  issn = {24058963},
  copyright = {https://www.elsevier.com/tdm/userlicense/1.0/},
  langid = {english}
}

@article{bouillonPharmacodynamicInteractionPropofol2004,
  title = {Pharmacodynamic Interaction between Propofol and Remifentanil Regarding Hypnosis, Tolerance of Laryngoscopy, {{Bispectral Index}}, and Electroencephalographic Approximate Entropy},
  author = {Bouillon, Thomas~W. and Bruhn, J{\"o}rgen and Radulescu, Lucian and Andresen, Corina and Shafer, Thomas~J. and Cohane, Carol and Shafer, Steven~L.},
  year = 2004,
  month = jun,
  journal = {Anesthesiology},
  volume = {100},
  number = {6},
  pages = {1353--1372},
  issn = {0003-3022},
  langid = {english}
}

@misc{raymondMPCTrackingAnesthesia2025,
  title = {{{MPC}} for Tracking for Anesthesia Dynamics},
  author = {Raymond, Maxim and Moussa, Kaouther and Fiacchini, Mirko and Lauber, Jimmy},
  year = 2025,
  month = dec,
  howpublished = {arXiv:2512.08452},
  number = {arXiv:2512.08452},
  eprint = {2512.08452},
  primaryclass = {eess},
  publisher = {arXiv},
  doi = {10.48550/arXiv.2512.08452},
  archiveprefix = {arXiv}
}

@article{eleveldAllometricModelRemifentanil2017,
  title = {An Allometric Model of Remifentanil Pharmacokinetics and Pharmacodynamics},
  author = {Eleveld, Douglas J. and Proost, Johannes H. and Vereecke, Hugo and Absalom, Anthony R. and Olofsen, Erik and Vuyk, Jaap and Struys, Michel M. R. F.},
  year = 2017,
  month = jun,
  journal = {Anesthesiology},
  volume = {126},
  number = {6},
  pages = {1005--1018},
  issn = {0003-3022},
  doi = {10.1097/ALN.0000000000001634},
  langid = {english}
}

@article{eleveldPharmacokineticPharmacodynamicModel2018,
  title = {Pharmacokinetic--Pharmacodynamic Model for Propofol for Broad Application in Anaesthesia and Sedation},
  author = {Eleveld, Douglas J. and Colin, P. and Absalom, Anthony R. and Struys, Michel M. R. F.},
  year = 2018,
  month = may,
  journal = {British Journal of Anaesthesia},
  volume = {120},
  number = {5},
  pages = {942--959},
  issn = {00070912},
  doi = {10.1016/j.bja.2018.01.018},
  copyright = {https://www.elsevier.com/tdm/userlicense/1.0/},
  langid = {english}
}

@article{aubouin-pairaultComparisonMultipleKalman2024,
  title = {Comparison of Multiple {{Kalman}} Filter and Moving Horizon Estimator for the Anesthesia Process},
  author = {{Aubouin-Pairault}, Bob and Fiacchini, Mirko and Dang, Thao},
  year = 2024,
  month = apr,
  journal = {Journal of Process Control},
  volume = {136},
  pages = {103179},
  issn = {09591524},
  doi = {10.1016/j.jprocont.2024.103179},
  langid = {english}
}

@article{aubouin-pairaultPASPythonAnesthesia2023,
  title = {{{PAS}}: A Python Anesthesia Simulator for Drug Control},
  shorttitle = {{{PAS}}},
  author = {{Aubouin-Pairault}, Bob and Fiacchini, Mirko and Dang, Thao},
  year = 2023,
  month = aug,
  journal = {Journal of Open Source Software},
  volume = {8},
  number = {88},
  pages = {5480},
  issn = {2475-9066},
  doi = {10.21105/joss.05480},
  langid = {english}
}

@article{mintoInfluenceAgeGender1997,
  title = {Influence of Age and Gender on the Pharmacokinetics and Pharmacodynamics of Remifentanil: {{I}}. Model Development},
  shorttitle = {Influence of {{Age}} and {{Gender}} on the {{Pharmacokinetics}} and {{Pharmacodynamics}} of {{Remifentanil}}},
  author = {Minto, Charles F. and Schnider, Thomas W. and Egan, Talmage D. and Youngs, Elizabeth and Lemmens, Harry J. M. and Gambus, Pedro L. and Billard, Valerie and Hoke, John F. and Moore, Katherine H. P. and Hermann, David J. and Muir, Keith T. and Mandema, Jaap W. and Shafer, Steven L.},
  year = 1997,
  month = jan,
  journal = {Anesthesiology},
  volume = {86},
  number = {1},
  pages = {10},
  issn = {1528-1175},
  doi = {10.1097/00000542-199701000-00004},
  langid = {american}
}

@article{schniderInfluenceAgePropofol1999,
  title = {The Influence of Age on Propofol Pharmacodynamics},
  author = {Schnider, Thomas W. and Minto, Charles F. and Shafer, Steven L. and Gambus, Pedro L. and Andresen, Corina and Goodale, David B. and Youngs, Elizabeth J.},
  year = 1999,
  month = jun,
  journal = {Anesthesiology},
  volume = {90},
  number = {6},
  pages = {1502},
  issn = {1528-1175},
  doi = {10.1097/00000542-199906000-00003},
  langid = {american}
}

@article{queinnecSwitchedControlStrategy2019,
  title = {Switched Control Strategy Including Time Optimal Control and Robust Dynamic Output Feedback for Anaesthesia},
  author = {Queinnec, Isabelle and Tarbouriech, Sophie and Zabi, Sa{\"i}d and Garcia, Germain and Mazerolles, Michel},
  year = 2019,
  journal = {IET Control Theory \& Applications},
  volume = {13},
  number = {7},
  pages = {960--969},
  issn = {1751-8652},
  doi = {10.1049/iet-cta.2018.5260},
  copyright = {\copyright{} 2021 The Authors. IET Control Theory \& Applications published by John Wiley \& Sons, Ltd. on behalf of The Institution of Engineering and Technology},
  langid = {english}
}

@article{ionescuNonlinearDynamicsPatients2015,
  title = {Nonlinear Dynamics of the Patient's Response to Drug Effect during General Anesthesia},
  author = {Ionescu, Clara and Tenreiro Machado, Jose and De Keyser, Robin and Decruyenaere, Johan and Struys, Michel M. R. F.},
  year = 2015,
  month = mar,
  journal = {Communications in Nonlinear Science and Numerical Simulation},
  volume = {20},
  number = {3},
  pages = {914--926},
  issn = {1007-5704},
  doi = {10.1016/j.cnsns.2014.05.014}
}

@book{bookMayneRawlings,
author = {Rawlings, J. and Mayne, D.Q. and Diehl, Moritz},
year = {2017},
month = {01},
pages = {},
title = {Model Predictive Control: Theory, Computation, and Design}
}

@inproceedings{ajamiModelPredictiveControl2026,
  title = {Model {{Predictive Control}} for Epidemic Containment},
  booktitle = {2026 34th {{Mediterranean Conference}} on {{Control}} and {{Automation}} ({{MED}})},
  author = {Ajami, Mohamad and Sonveaux, Candy and Dumont, Morgane and Fiacchini, Mirko},
  year = 2026,
  month = jun,
  pages = {159--164},
  issn = {2473-3504},
  doi = {10.1109/MED70602.2026.11598367}
}

@article{BEMPORAD20023,
title = {The explicit linear quadratic regulator for constrained systems},
journal = {Automatica},
volume = {38},
number = {1},
pages = {3-20},
year = {2002},
issn = {0005-1098},
doi = {https://doi.org/10.1016/S0005-1098(01)00174-1},
url = {https://www.sciencedirect.com/science/article/pii/S0005109801001741},
author = {Alberto Bemporad and Manfred Morari and Vivek Dua and Efstratios N. Pistikopoulos}
}
\bibliographystyle{plain}

\end{document}